\documentclass[aps,prab,preprint,superscriptaddress]{revtex4-2}
\usepackage{graphicx}
\usepackage{comment}
\usepackage{nameref}
\begin{document}

%\preprint{}

%\title{A beam extraction scheme for the surface muon beamline at SHINE}
\title{Simulation studies of a high-repetition-rate electron-driven surface muon beamline at SHINE}


%\author{}
%\thanks{}
%\affiliation{}

\author{Fangchao Liu}
\thanks{These authors contributed equally to this work}
\author{Yusuke Takeuchi}
\thanks{These authors contributed equally to this work}
\affiliation{Tsung-Dao Lee Institute and School of Physics and Astronomy, Shanghai Jiao Tong University, Shanghai 201210, China}
\author{Si Chen}
\affiliation{Shanghai Advanced Research Institute, Chinese Academy of Sciences, Shanghai 201210, China}
\author{Siyuan Chen}
\author{Kim Siang Khaw}
\email{kimsiang84@sjtu.edu.cn}
\author{Meng Lyu}
\affiliation{Tsung-Dao Lee Institute and School of Physics and Astronomy, Shanghai Jiao Tong University, Shanghai 201210, China}
\author{Ziwen Pan}
\affiliation{State Key Laboratory of Particle Detection and Electronics, University of Science and Technology of China, Hefei 230026, China}
\author{Dong Wang}
\affiliation{Shanghai Advanced Research Institute, Chinese Academy of Sciences, Shanghai 201210, China}
\author{Jiangtao Wang}
\author{Liang Wang}
\affiliation{Tsung-Dao Lee Institute and School of Physics and Astronomy, Shanghai Jiao Tong University, Shanghai 201210, China}
\author{Wenzhen Xu}
\affiliation{Shanghai Advanced Research Institute, Chinese Academy of Sciences, Shanghai 201210, China}

\date{\today}

\begin{abstract}
A high-repetition-rate pulsed muon source operating at approximately 50\,kHz holds the potential to improve the sensitivity of various particle physics and material science experiments involving muons. In this article, we propose utilizing the high-repetition-rate pulsed electron beam at the SHINE facility to generate a surface muon beam. Our simulation studies indicate that an 8\,GeV, 100\,pC charge pulsed electron beam impinging on a copper target can produce up to $2 \times 10^{3}$ muons per pulse. Beamline optimization results demonstrate that approximately 60 surface muons per electron bunch can be efficiently transported to the end of the beamline. This translates to a surface muon rate of $3 \times 10^{6}\,\mu^{+}$/s when the pulsed electron beam is operated at 50\,kHz, which is comparable to existing muon facilities. This high-repetition-rate pulsed muon beam, with its ideal time structure, represents a unique and pioneering effort once constructed. It serves as a model for building cost-effective muon sources at existing electron machines with GeV electron energies. In addition to the typical challenges encountered in conventional muon beamlines, such as the installation and construction of the target station and beamline, the removal of substantial quantities of positrons is also a major challenge. A potential solution to this issue is also discussed.
\end{abstract}

\maketitle

\section{Introduction}
Muons play a pivotal role in various scientific domains, including particle physics, nuclear physics, and condensed matter physics~\cite{Gorringe:2015cma,Hillier:2022nat}, due to their unique properties and interactions. The increasing importance and demand for muons in these fields have led to a growing emphasis on their production and application. Today, muons can be generated in large quantities by bombarding target materials with intense proton beams from advanced accelerator facilities. This process primarily involves the production of pions through strong nuclear interactions, which then decay into muons.

Of particular interest are ``surface muons'', which originate from pions decaying at rest near the target surface. These muons exhibit nearly 100\% polarization due to the parity-violating weak decay of pions, and their momentum remains almost monochromatic at approximately 29.8\,MeV/c. These distinctive properties make surface muons highly valuable for a wide range of experimental studies. Current state-of-the-art muon facilities, such as PSI~\cite{Grillenberger:2021kyv}, RCNP~\cite{Cook:2016sfz} and J-PARC~\cite{Kawamura:2018apy}, deliver usable surface muon beams with intensities ranging from $10^7$ to $10^8\,\mu^{+}$/s to the experimental area of muon spin rotation ($\mu$SR) experiments.

\begin{figure}[htbp]
\includegraphics[width=0.8\linewidth]{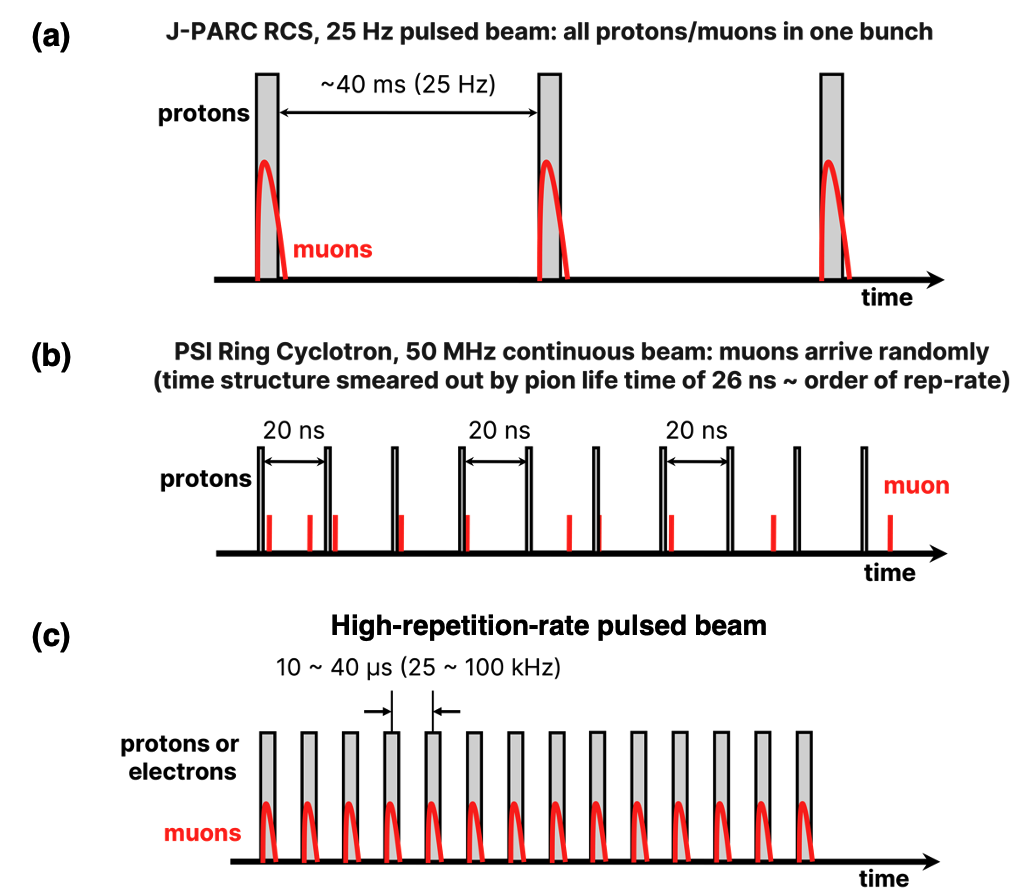}
\caption{Typical time structure of (a) pulsed and (b) continuous muon sources. (c) An ideal pulsed muon source with a high repetition rate. Adapted from~\cite{Cywinski:2009zz}.}
\label{fig:beam_time_structure}
\end{figure}

However, the repetition rates of muon beams currently available at these facilities are non-ideal and essentially limited to two modes: pulsed mode (25-50 Hz, e.g., J-PARC) and continuous mode (e.g., PSI). Considering that a typical muon experiment lasts for about ten muon lifetimes, current operation modes are not ideal for those muon experiments~\cite{Cywinski:2009zz,Adelmann:2010zz,Willmann:2021boq}. For example, the duty cycle is low in the pulsed mode, and the continuous mode is disadvantageous in terms of statistics. The typical time structure of currently available muon beams is shown in Fig.\ref{fig:beam_time_structure}. Many authors have mentioned that the most suitable muon source for experiments — such as $\mu$SR~\cite{Cywinski:2009zz}, muon electric dipole moment~\cite{Adelmann:2010zz}, muonium to anti-muonium conversion~\cite{Willmann:2021boq}, muon lifetime~\cite{Kanda:2022too}, and muon spin force~\cite{Ema:2023pac} — operates in a pulsed mode with a repetition rate of several tens of kHz, as shown in Fig.~\ref{fig:beam_time_structure}(c).

To meet this demand, improvements to the time structure in muon production using existing proton beams are being considered.
There are several plans to generate proton beams with a time structure suitable for muon sources with ideal repetition rates. 
For instance, the development of a non-scaling fixed-field alternating gradient (FFAG) proton accelerator operating at frequencies of a few kHz has been proposed~\cite{Kuno:2000wn,Seidel:2021jyz}; in the high-energy beam transport of a spallation neutron source (SNS) accelerator, proton pulses with a repetition rate of 50 kHz have been successfully extracted based on laser neutralization of a 1\,GeV hydrogen ion (H$^{-}$) beam~\cite{Liu:2020hcu}; the proton beam for the Mu2e experiment at Fermilab has a proton bunch repetition rate of 0.59\,MHz, which is achieved by resonant slow extraction of the proton bunches from the delivery ring~\cite{Mu2e:2014fns}. However, these attempts have not yet been realized or are not versatile enough.

On the other hand, muon sources utilizing electron beams have recently emerged as a promising alternative to traditional proton-driven sources. Nagamine notably proposed the use of electron microtrons for a compact $\mu$SR beamline~\cite{Nagamine:2009zz}. Recent advancements have established high-repetition-rate (kHz to MHz) electron beams generated by linear accelerators. When combined with a muon production target, these electron beams can create muon sources with an optimal time structure.

Unlike proton beam-driven sources, electron beam-driven schemes produce muons as tertiary beams through photo-nuclear processes. In this method, Bremsstrahlung generates real photons, which subsequently induce photo-excitation of nuclei, leading to pion production and subsequently decay muons. Additionally, muons can be produced via the Bethe-Heitler pair production process, although this method has a lower production cross-section. Despite this, the Bethe-Heitler process yields higher-energy muons with better directionality, as it bypasses pion production and decay stages.

This concept is particularly attractive because it does not require a dedicated facility and can be implemented at any facility that produces electron beams for various research applications. It is especially compatible with synchrotron radiation and X-ray free electron laser (XFEL) facilities, where GeV electron beams are typically dumped, thus offering a sustainable and innovative approach to muon production.

Furthermore, recent advances in laser wakefield acceleration (LWFA) technology have reduced the size of electron accelerators from kilometers to meters, further driving research into compact muon source concepts. The anticipated availability of high-repetition-rate femtosecond multi-PW lasers has fueled this research momentum. However, most studies utilizing laser technology~\cite{Titov:2009cr,Dreesen:2014mt,Rao:2018njj,Calvin:2023, Zhang:2024axy, Terzani:2024neq} have focused on generating high-energy muons through the pair production process rather than low-energy muons for $\mu$SR applications.

In this article, we explore the use of high-repetition-rate electron beams from the "Shanghai High repetition rate XFEL and Extreme Light facility" (SHINE)~\cite{Zhao:2018lcl} as a driver for muon sources. This cutting-edge facility, currently under construction in Zhangjiang, Shanghai, will feature a continuous-wave (CW) superconducting electron linac capable of delivering an 8-GeV bunched electron beam with a bunch charge of 100\,pC and a repetition rate of up to 1\,MHz, resulting in an average current of 100\,$\mu$A. The beamline will include three undulator lines capable of generating hard X-rays up to 25\,keV, and a fast kicker system will distribute electron bunches, supplied at up to 1\,MHz, to the respective beamlines and beam dumps. More information about the current layout can be found at~\cite{Liu:2024fcg}.

\begin{figure*}[htbp]
\includegraphics[width=0.95\linewidth]{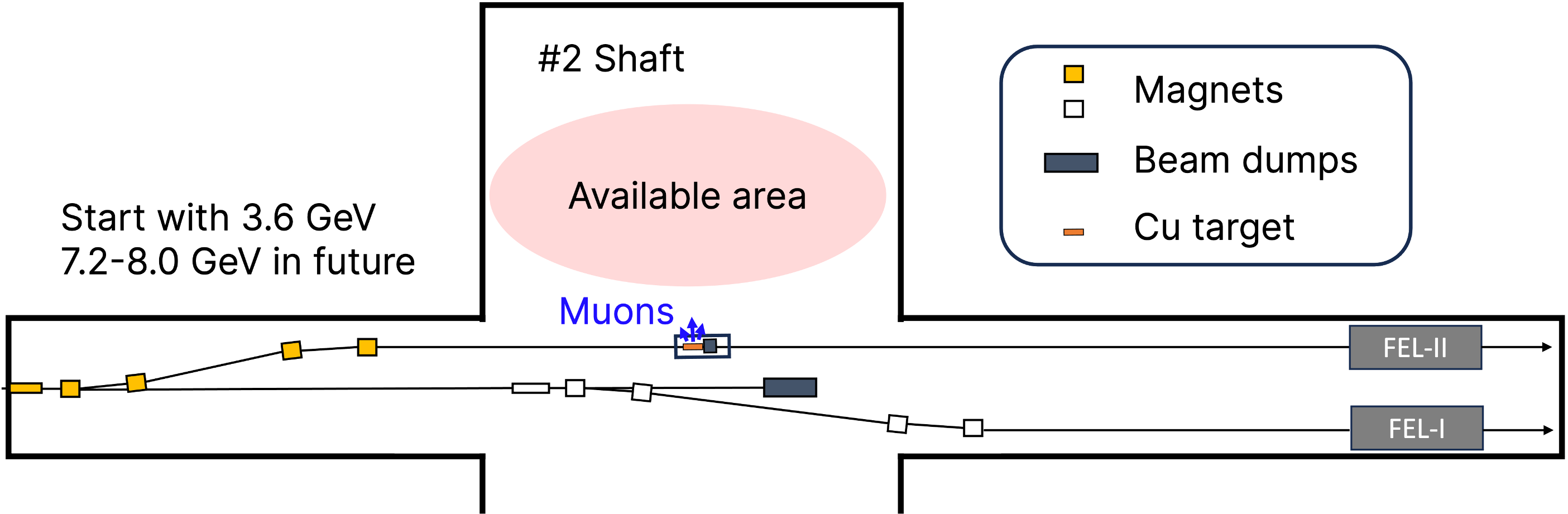}
\caption{A conceptual schematic of the SHINE muon source project. A large number of 8 GeV electron beam bunches fed from the superconducting accelerator at repetition rates up to 1\,MHz are distributed by a fast kicker in the beam switch yard to three FEL beamlines and several beam dumps. The beam going to the beam dump is irradiated to the muon production target, which is a combination of target and dump, to produce muons. The target can be retracted from the beam axis when not in use, so it does not interfere with FEL-II operation.}
\label{fig:shine_schematic}
\end{figure*}

The interaction between the electron beam and either the beam dump or a pre-beam dump thin target holds the potential for producing muons and other secondary particles~\cite{Lv:2023cmh}. Figure~\ref{fig:shine_schematic} illustrates the configuration of the electron beamline at the SHINE facility and provides a conceptual diagram of the muon source using this beam.

This paper outlines the concept of a high-repetition-rate muon source and a transport beamline for the muon beam, utilizing the high-repetition electron beam from the SHINE facility. Additionally, we discuss the expected surface muon beam intensity and its potential applications.

\section{Muon production target}
\subsection{Material selection}

For the production target material, we compared graphite, which is widely used in proton-driven muon sources, with copper and tungsten, as suggested in previous studies, to evaluate their effectiveness in muon production using an electron beam. We also included other common target materials such as Aluminum, Titanium, Iron, and Molybdenum to cover representative materials from low-$Z$, medium-$Z$, and high-$Z$ regions. Bremsstrahlung and pair production induce electromagnetic showers, which significantly influence muon production by the electron beam. Therefore, understanding the properties of these showers is critical. The characteristics of electromagnetic showers in different target materials, based on SHINE's design parameters (8\,GeV, 100\,$\mu$A), are summarized in Table~\ref{tab:mat_properties}. We evaluated seven representative materials based on key properties: the depth of the shower maximum ($X_{\rm{max}}$), the heat load ($dE/dz$) at this depth, and typical thermal conductivity ($\kappa$).

\begin{table}[htbp]
\centering
\caption{The electromagnetic shower characteristics for 8\,GeV, 100\,$\mu$A beam and material properties. For heat loads, those at 50\,kHz operation are also shown.}
\label{tab:mat_properties}
\begin{tabular}{lcccccccc}
\toprule
 Materials & C (Graphite) & Al & Ti & Fe & Cu & Mo & W\\
\hline
$Z$ & 6 & 13 & 22 & 26 & 29 & 42 & 74 \\
$A$ & 12 & 27 & 48 & 56 & 64 & 96 & 184   \\
$\rho$ [g/cm$^{3}$] & 1.82 & 2.70 & 4.54 & 7.87 & 8.94 & 10.22 & 19.25 \\
$X_{\rm{max}}$ [mm] & 958 & 425 & 191 & 96.6 & 80.5 & 57.1 & 22.4 \\
$dE/dz$ [kW/cm] (1\,MHz) & 4.84  & 11.6   & 27.3   & 53.4   & 65.4   & 71.9   & 265   \\
$dE/dz$ [kW/cm] (50\,kHz) & 0.24  & 0.58   & 1.37   & 2.67   & 3.27   & 3.60   & 13.3 \\
$\kappa$ [W\,m$^{-1}$\,K$^{-1}$] & 119 & 237 & 21.9 & 80.4 & 401 & 138 & 173   \\
\hline
\end{tabular}
\end{table}

\begin{figure}[htbp]  
    \includegraphics[width=0.8\linewidth]{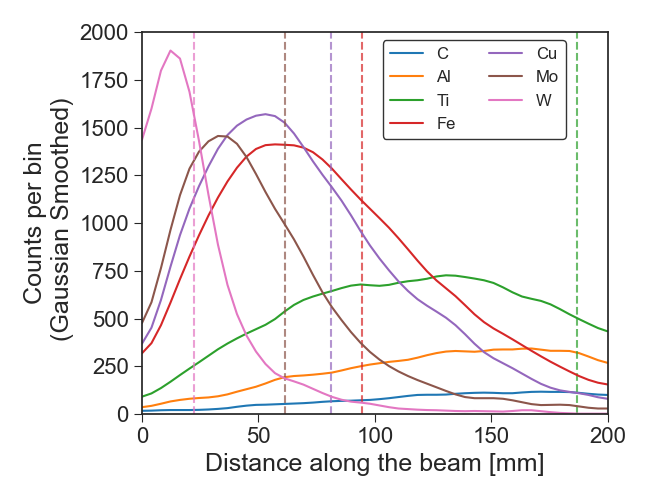}  
    \caption{Pion stopping distribution along the beam direction and calculated shower maximum $X_{\rm{max}}$ (dashed lines) for different materials. Data are binned in 4\,mm intervals and smoothed. }  
    \label{fig:pion_longitudinal_distribution}  
\end{figure}

\begin{figure}[htbp]  
    \includegraphics[width=0.8\linewidth]{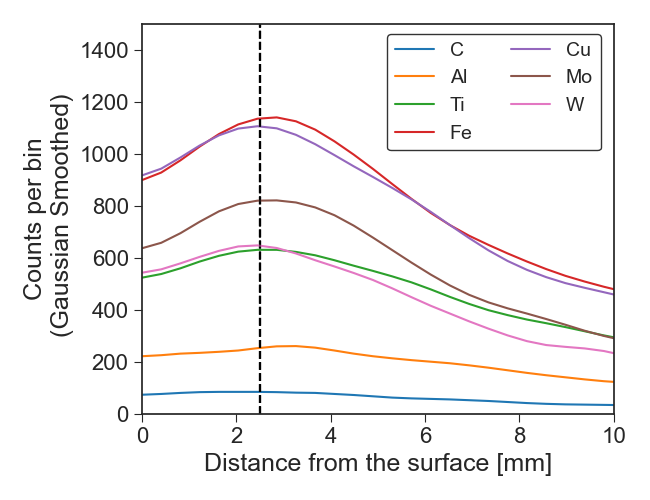}  
    \caption{Pion stopping distribution transverse to the beam for different materials, with the beam position (dashed line) at 2.5\,mm from the target surface. Data are binned in 0.4\,mm intervals and smoothed. }  
    \label{fig:pion_transverse_distribution}  
\end{figure}

In particular, $X_{\rm{max}}$ plays a key role in determining the pion production and stopping distributions. As shown in Fig.~\ref{fig:pion_longitudinal_distribution}, the peak of the pion stopping distribution occurs upstream of the calculated shower maximum $X_{\rm{max}}$, likely because, as the electromagnetic shower develops, the average secondary-particle energy drops below the threshold required for $\Delta$-resonance excitation, which mainly contributes to pion production. Consequently, medium-Z materials (e.g., Fe, Cu) exhibit optimal characteristics, as they efficiently stop pions within the target and maximize the number of pions coming to rest near the surface ($z = 2.5$), which is crucial for surface muon production, as illustrated in Fig.~\ref{fig:pion_transverse_distribution}. For each dataset, simulations were carried out in G4Beamline~\cite{Roberts:2007nte} (physics model QGSP‑BERT), irradiating the target with 10\,million 8\,GeV electrons. The simulation configuration is the same as in Fig.~\ref{fig:shine_muon_target}.

Due to its energy absorption characteristics, low-$Z$ materials necessitate an extended length for complete electromagnetic shower development. This increased target dimension is suboptimal for surface muon production, as the initial beam size correlates directly with target dimensions. While high-$Z$ materials enable electromagnetic shower development over shorter distances, it presents significant thermal management challenges due to concentrated heat loads in targets. Medium-$Z$ materials emerge as the optimal balance, facilitating electromagnetic shower development over moderate lengths while generating reasonably lower heat load. Furthermore, copper's superior thermal conductivity characteristics make it particularly suitable for managing the thermal stress induced by high-repetition, high-energy electron beams. The material's well-established manufacturing processes and cost-effectiveness, combined with these favorable physical properties, position copper as the prime candidate for high-repetition muon production targets, despite certain remaining technical challenges. 

Based on these considerations, we will proceed with copper as the target material for further studies.

\subsection{Target geometry}

The thin slab targets used at PSI~\cite{Berg:2015wna} and in the CSNS target design studies~\cite{Chen:2023uvp} are considered standard for proton-based muon sources. These facilities must maintain a material budget in the proton beam and scale the target length accordingly due to downstream spallation neutron sources. Occasionally, targets need to be tilted slightly (approximately 5\,degrees) to increase the target size while maintaining the interaction length. Thin targets pose challenges in heat management, sometimes necessitating rotating targets, which can introduce additional mechanical issues.

In contrast, our muon source utilizes an electron beam that is dumped directly, eliminating the need for a thin target to maintain the material budget. To ensure radiation safety beyond the tunnel's limited shielding thickness, it is essential to provide comprehensive local shielding around the high-power target. This configuration allows the use of a box-shaped thick target and a beam dump without significant disadvantages, facilitating cooling by increasing the target's heat capacity.

\begin{figure}[htbp]
\includegraphics[width=0.9\linewidth]{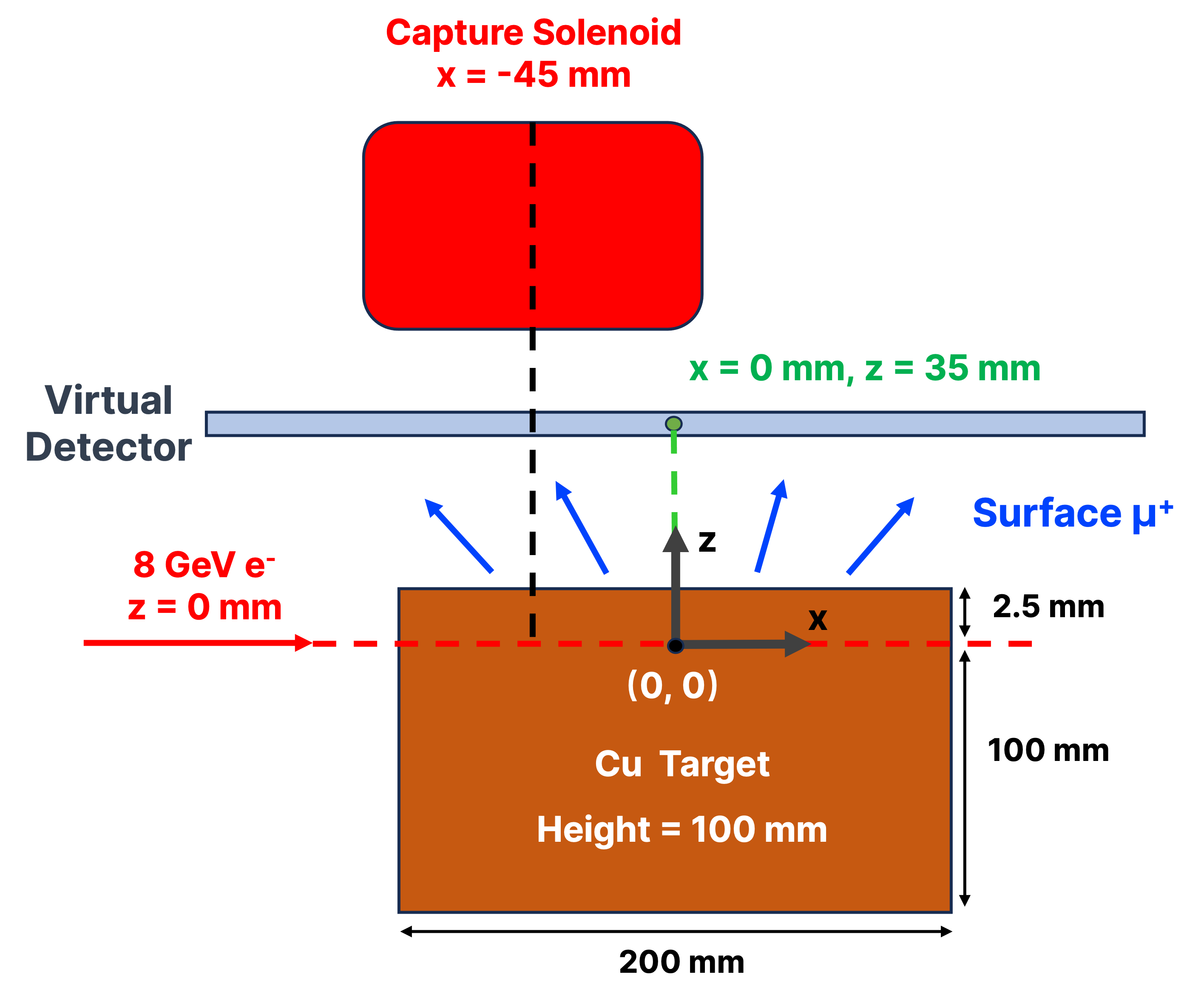}
\caption{The schematic view of the target at SHINE surface muon source. In our setup, the 8\,GeV electron beams impinging on the thick copper target along the positive direction of the $x$-axis, and a virtual detector is placed to monitor muon beam information at the positive half of the $z$-axis, 35\,mm from the electron beam. The centroid of the muon beam falls on the central axis of the solenoid.}
\label{fig:shine_muon_target}
\end{figure}

The beam position and target length were optimized to maximize the surface muon yield. Particle interactions and transport processes in the target were simulated using FLUKA code, version 4-4.0~\cite{Ferrari:2005zk} to estimate the yield. The simulation assumed an 8\,GeV, 100\,pC/bunch electron beam with an RMS beam size of 2\,mm impinged on the target~\footnote{The the typical electron beam spot from the SHINE superconducting linac is about $\mathcal{O}$(10 $\mu m$). However, at this small beam size, it will produce excessively high power density. At a repetition rate of 50\,kHz, the copper target may not endure the thermal load. The 2\,mm beam spot assumed here is based on a recent SHINE beam dump study~\cite{Xu:2020}, and this number will be optimized with a more focused study based on the target's thermal engineering design. To achieve a millimeter-sized beam spot, a beam spoiler system can be utilized. Such a spoiler system is commonly employed at the end of the SLAC linac to degrade the beam at the A-line and ESA~\cite{Nosochkov:2019ehm}.}. A virtual detector, measuring 100\,cm $\times$ 100\,cm, was positioned 35\,mm perpendicular to the electron beam to record muon counts. The target configurations in the simulation are shown in Fig.~\ref{fig:shine_muon_target}. The target simulations were conducted without a capture solenoid, and the capture solenoid's position was later determined to align with the beam center, as detailed in Section III.B.

A variance reduction technique, specifically mean-free path biasing (LAM-BIAS), was introduced into the computational model to improve efficiency. An energy threshold of 10\,MeV was set to ignore the production and transport of low-energy electrons, positrons, and photons, which do not contribute to muon production, enhancing calculation speed. For each geometry, $3.0 \times 10^8$ electrons were simulated to balance statistical reliability with computational cost, with the biasing method shortening the photon hadronic interaction length by a factor of 0.02. The results were then scaled to match the SHINE bunch charge of $6.25 \times 10^8$ electrons per bunch.

\begin{figure}[htbp]  
    \includegraphics[width=0.9\linewidth]{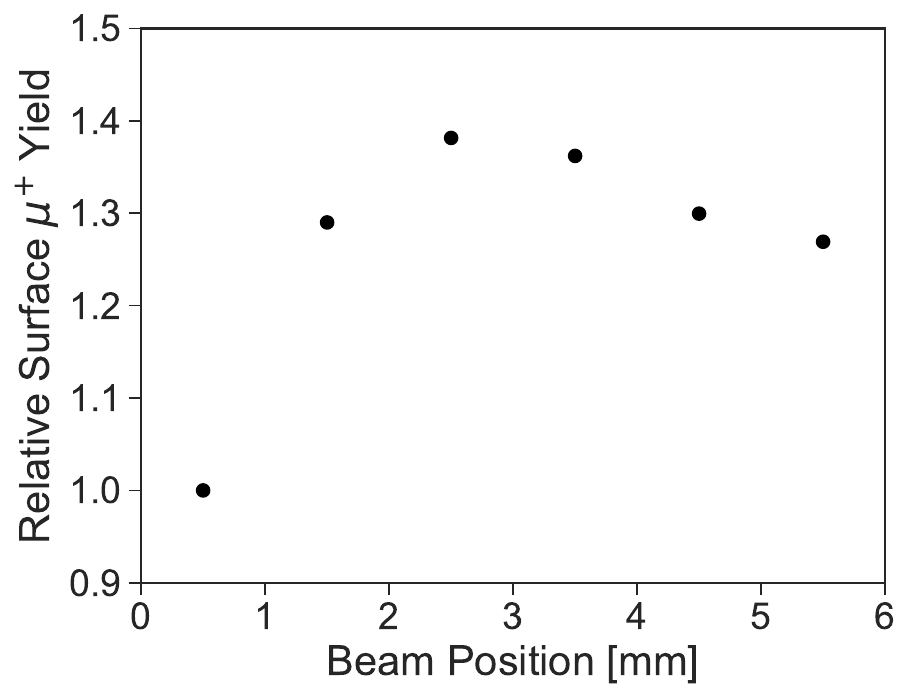}  
    \caption{Dependence of surface muon yield on electron beam spot position. Among the muons produced by photo-nuclear process detected by the virtual detector, those with a momentum of 25 to 30 MeV/c were selected.}  
    \label{fig:electron_beam_position}  
    \end{figure}

\begin{figure}[htbp]  
    \includegraphics[width=0.9\linewidth]{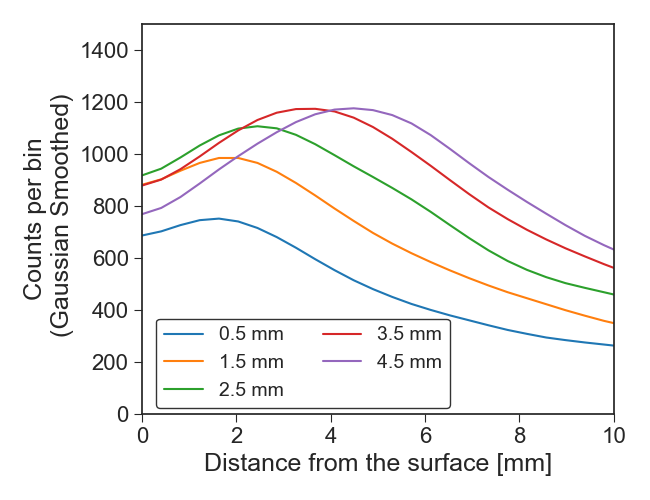}  
    \caption{Beam position dependence of pion distribution in the target. Each line corresponds to the results for beam positions of 0.5, 1.5, 2.5, 3.5, and 4.5\,mm. Simulation and plot settings are the same as in Fig.~\ref{fig:pion_transverse_distribution}.}  
    \label{fig:beam_position_scan_for_pion}  
    \end{figure}

The beam position was varied from 0.5 to 5.5\,mm away from the target surface in 1\,mm increments. Figure~\ref{fig:electron_beam_position} shows the surface muon yield at each point normalized by the yield at 0.5\,mm. The yield increased rapidly until the beam position reached 2.5\,mm, beyond which it gradually decreased. Further investigation of the pion stopping distribution across the target would help explain this relationship more clearly. As shown in Fig.~\ref{fig:beam_position_scan_for_pion}, when the beam position moves farther from the surface, the peak value of the pion distribution within the target reaches its maximum at 3.5\,mm, due to the increased volume available for shower lateral development. However, the number of pion stops near the surface which is crucial for surface muon production, begins to decrease beyond 2.5\,mm as the pion distribution shifts away from the surface. The interplay of these two effects indicates that 2.5\,mm is the optimal position.

\begin{figure}[htbp]  
    \includegraphics[width=0.9\linewidth]{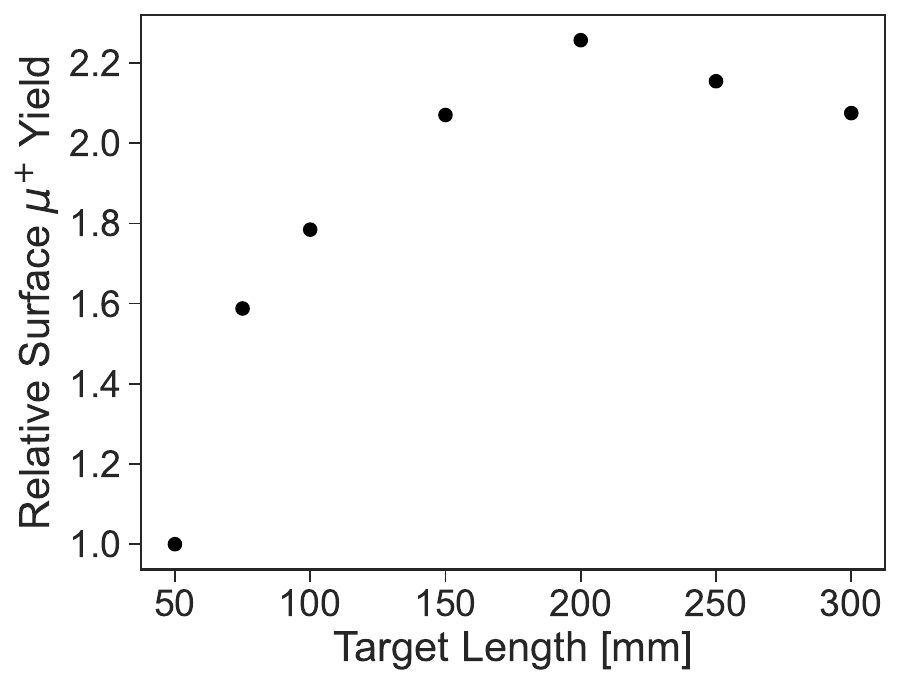}  
    \caption{Dependence of surface muon yield on target length. Among the muons produced by the photo-nuclear process detected by the virtual detector, those with a momentum of 25 to 30 MeV/c were selected.}  
    \label{fig:target_length_scan}  
\end{figure}

With the beam position fixed at 2.5\,mm, the target length along the beam direction was scanned, examining seven points between 50\,mm and 300\,mm. As shown in Fig.~\ref{fig:target_length_scan}, surface muon yield increased up to a target length of 200\,mm before decreasing. Notably, the enhancement from 50 to 75\,mm was substantial. As shown in Tab.~\ref{tab:mat_properties}, $X_{\rm max}$ is 80.5\,mm, suggesting significant benefits from increasing the target length up to this point. 
As shown in Fig.~\ref{fig:pion_longitudinal_distribution}, the pion stopping distribution exhibits a exponential decrease up to around 200\,mm, extending the target length can further increase yield, with a 200\,mm target length set as our baseline based on these findings.

\section{Surface muon beamline}
To optimize the muon production target parameters, we employed FLUKA for its high computational efficiency. Since FLUKA is not designed for detailed particle tracking simulation, we used G4Beamline—a code specializing in particle transport and optics—to generate initial particle distributions for beamline design studies. This approach allowed us to seamlessly transition to subsequent G4Beamline studies. 

\subsection{Surface muon yield}
\begin{figure}[htbp]
\includegraphics[width=0.8\linewidth,clip]{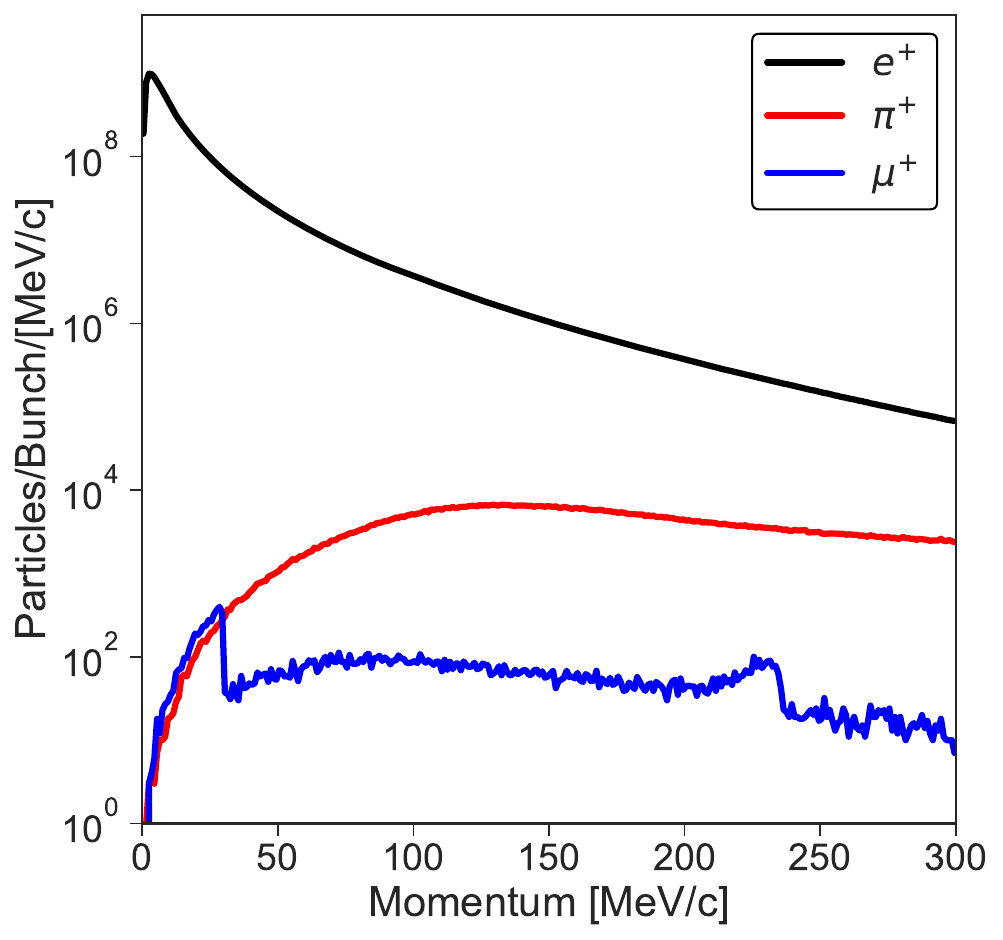}
\caption{Momentum distribution of the detected secondary and tertiary particles at the virtual detector.}
\label{fig:momentum_distribution_for_secondaries}
\end{figure}

Firstly, we thoroughly investigated the surface muon yield using optimized baseline target designs with G4Beamline. The physics model used was QGSP-BERT with a simulation setup akin to the FLUKA setup (a virtual detector is placed near the side of the target). The yield of positive muons (below 300\,MeV/c) from photo-nuclear processes was estimated at approximately $10^{4}$ per bunch. Fig.~\ref{fig:momentum_distribution_for_secondaries} shows the momentum distribution of these photo-nuclear-generated muons along with other major positively charged particles. Comparing muon yields between the G4Beamline and the FLUKA simulations showed agreement within 2\% for momenta below 300 MeV/c, and within 30\% for the 25-30 MeV/c momentum region (see details in~\appendixname).

Muons exhibit a broad energy distribution, decreasing from low to high energy regions. Notably, the momentum distribution shows two distinct peaks, corresponding to the production of surface muons from the decay of pions and kaons, with momentum peaks around 30\,MeV/c and 230\,MeV/c, respectively. The yield of surface muons derived from pion decay in the 25 to 30\,MeV/c range is approximately $2 \times 10^{3}$ muons per bunch. Operating at a frequency of 50\,kHz, this translates to an intensity of $1 \times 10^{8}\,\mu^{+}$/s. Although the muon yield per bunch is relatively low, the high repetition rate compensates for this, resulting in an overall intensity per second comparable to existing proton beam-driven muon sources\cite{Grillenberger:2021kyv,Cook:2016sfz,Kawamura:2018apy}.

Surface muons produced from kaon decays offer a unique opportunity to investigate extremely dense materials or materials under hydrostatic pressure due to their high penetrating power, allowing for deeper examination of such materials~\cite{Khasanov:2022hbd}.

In addition to muon-related experiments, pulsed positrons can be collected and transported to various dedicated terminals for applications in positron annihilation spectroscopy (PAS), which are associated with microstructural defects in materials~\cite{Tuomisto:2013,Selim:2021}. These terminals include positron lifetime measurements, Doppler broadening measurements, age-momentum correlation measurements, and angular correlation measurements. The exceptionally high intensity of the positrons can be effectively utilized to generate slow positron beams with adjustable energy, thereby meeting the significant demands in defect characterization within thin films and surface materials interfaces.

\subsection{Characteristics of the Initial Muon Beam}

\begin{figure*}[htbp]
\includegraphics[width=0.85\linewidth]{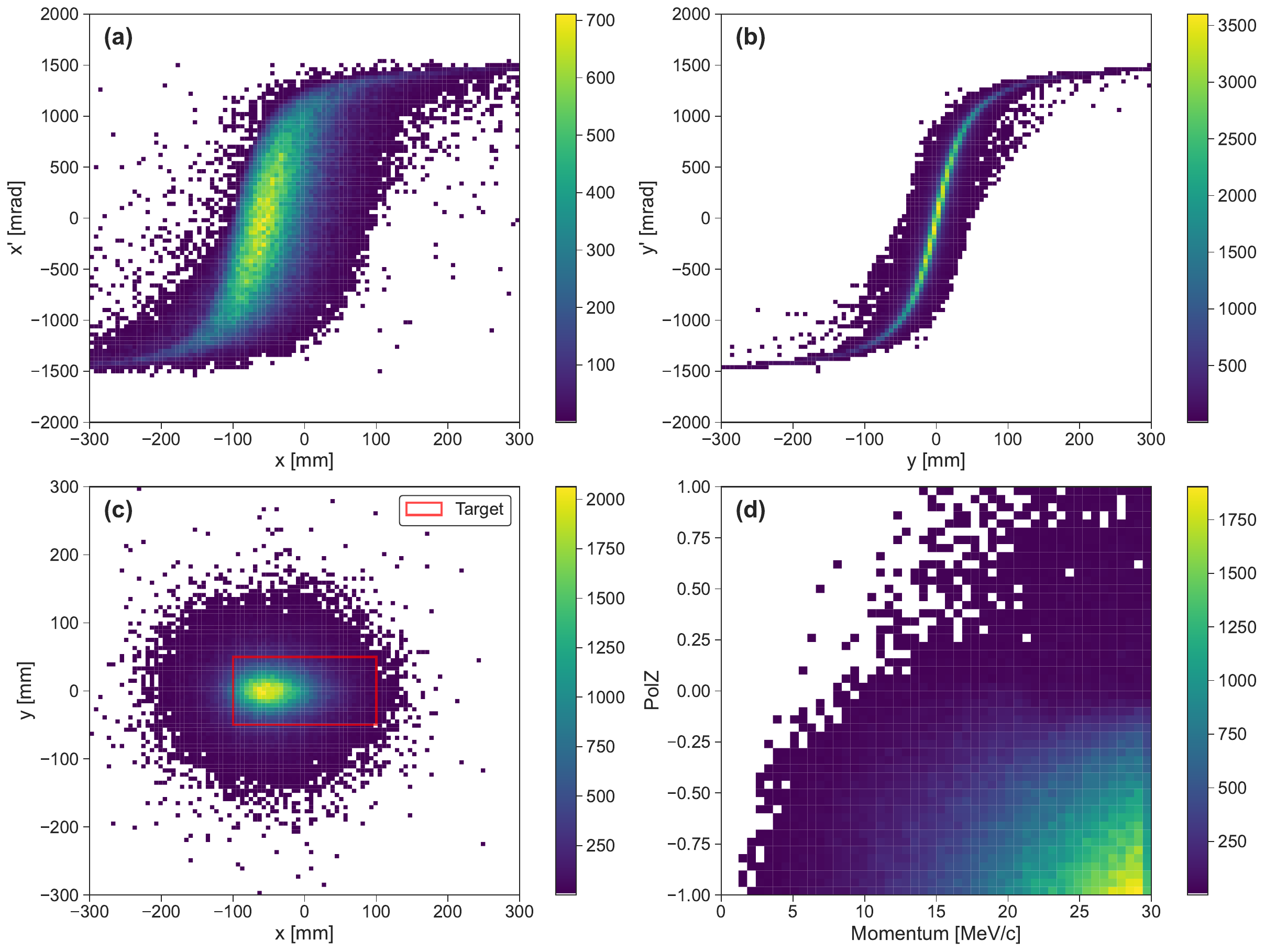}
\caption{(a) Horizontal phase space at the target ($x,x'$). (b) Vertical phase space at the target ($y,y'$). (c) Beam distribution at the target ($x,y$). (d) Momentum dependence of muon polarization, showing that high-momentum, forward-going muons exhibit maximum polarization. The coordinate system is the left-hand system in Fig.~\ref{fig:shine_muon_target}.}
\label{fig:beam_characteristic}
\end{figure*}

The characteristics of the muon beam emerging from the target were then analyzed using the virtual detector. Figure~\ref{fig:beam_characteristic} illustrates the horizontal and vertical phase space ($x-x'$, $y-y'$), the real space ($x-y$) of the beam, and the polarization dependence on momentum. From Fig.~\ref{fig:beam_characteristic} (Bottom right), it is evident that the muons with the highest polarization relative to the centerline-z axis possess the highest momentum, as these high-momentum muons are predominantly forward-going and abundant.

Selected phase space parameters of the muon beam are summarized in Tab.~\ref{tab:beam_source_param}. Given that the incident electron beam is aligned along the $x$-axis, the muon beam exhibits asymmetry in the horizontal plane, with a mean horizontal position shift of approximately -45.2\,mm from the center of the target. To maximize acceptance, the solenoid's cross-sectional center is aligned with the centroid of the beam.

\begin{table*}
\caption{Phase space characteristics of the muon beam detected in the simulation by the virtual detector, placed centrally 35\,mm at a 90-degree angle with respect to the electron beam.}
\label{tab:beam_source_param}
\begin{tabular} {lcccc}
\toprule
 Parameters & Value\\
\hline
Mean Momentum $P_0$ / Momentum Spread $\sigma_P$ & 22.9 MeV/c / 5.5  MeV/c\\
Horizontal Position $<x>$/ Width $\sigma_x$ & -45.2 mm / 73.7 mm \\
Horizontal Divergence $<x'>$/ Width $\sigma_{x^{'}}$ & -14.6 mrad / 696.7 mrad \\
Horizontal RMS Emittance  & 3814 $\pi$ cm mrad \\
Vertical Position $<y>$/ Width $\sigma_y$ & -0.1 mm / 67.5 mm  \\
Vertical Divergence $<y'>$/ Width $\sigma_{y^{'}}$ & -1.2 mm / 696.9 mrad \\
Vertical RMS Emittance  & 3180 $\pi$ cm mrad \\
Mean Polarization  &  -0.63   \\
\hline
\end{tabular}
\end{table*}

\subsection{Design concept of the beamline}

Beamlines are essential for collecting produced muons and selectively transporting muons with specific energies and properties to the experimental area, tailored for various scientific applications. While a straight, short beamline would offer the highest transmission rate, radiation safety concerns render such a layout impractical. Consequently, the beamline must incorporate several bending sections to eliminate direct line-of-sight from the experimental area to the target, also helping to minimize contamination by particles other than muons.

To streamline the optimization process, the beam optics design employs only solenoids and bending magnets, inspired by the beamline designs from PSI~\cite{Maso:2023zjp} and CSNS~\cite{Chen:2023uvp}. Despite the use of multiple bending magnets, contamination by positively charged particles with the same momentum as surface muons remains a possibility. This issue can be effectively mitigated with a Wien filter.

Based on the current layout design of the SHINE facility, the available space for the beamline and experimental area is situated in the north-central region of Shaft 2. The dimensions of this space are 30\,m $\times$ 12\,m, which must be considered in the design. Given these constraints, the beamline optics should be optimized to maximize the intensity of surface muons transported to the experimental area, ensuring both safety and efficiency.

\subsection{Beamline optics}

\begin{figure}[htbp]
\includegraphics[width=\linewidth]{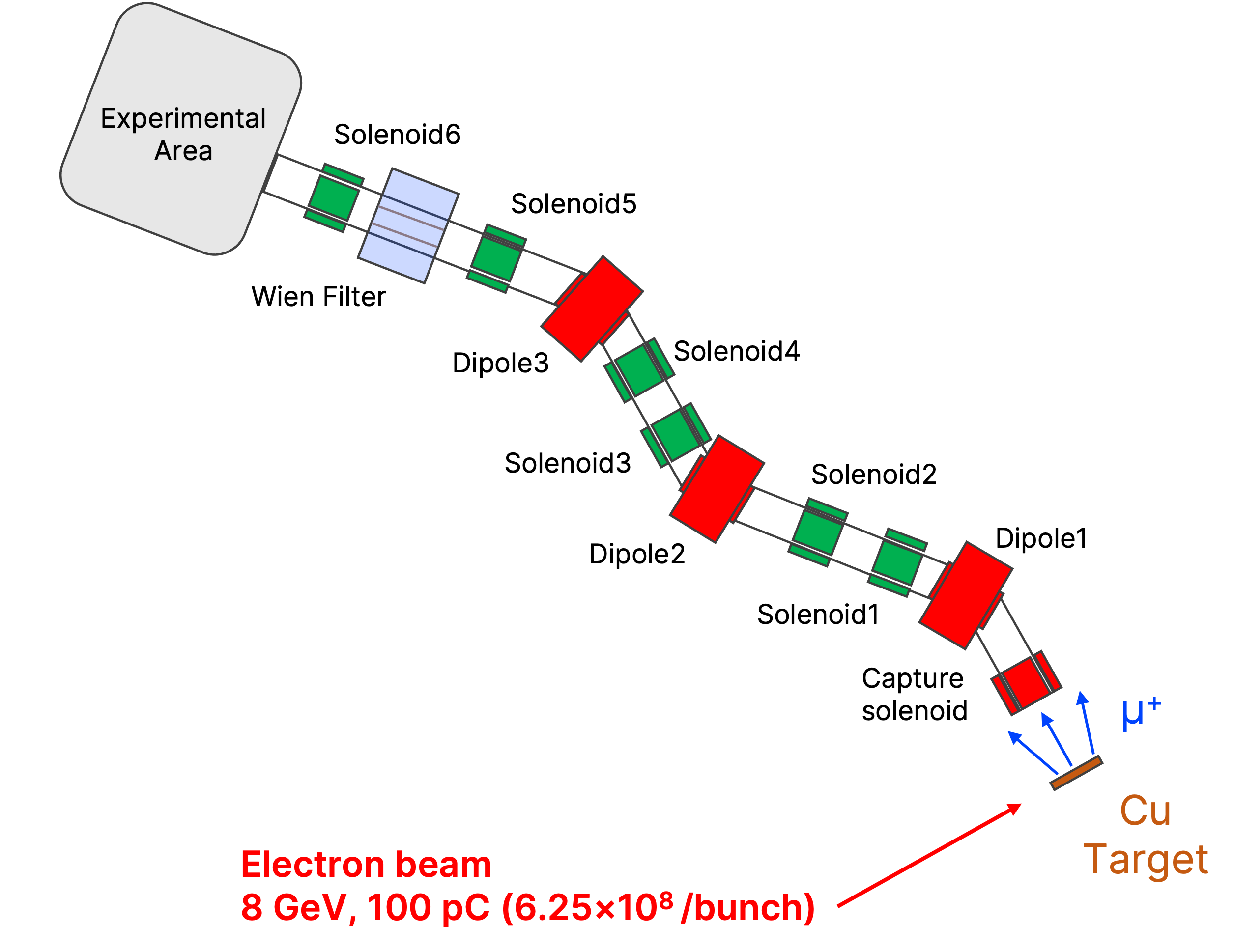}
\caption{The surface muon beamline layout.}
\label{fig:beamline_layout}
\end{figure}

Based on these design concepts, a 13.6-meter beamline, as shown in Fig.~\ref{fig:beamline_layout}, was constructed in the simulation to transmit surface muons from the target to the final experimental area. This beamline includes seven large-aperture solenoids (one capture solenoid and six focusing solenoids) and three dipole magnets with a 40-degree bending angle. Each solenoid is 373\,mm long with a 500\,mm aperture to ensure high acceptance.

To simplify the design, three 640-mm rectangular bending dipoles of identical geometry were used for momentum selection and the elimination of negatively charged and neutral particles, instead of sector dipoles. This approach allows for adjustments to the bending angle by simply changing the magnetic field inside the dipole, without the need for redesign. A Wien filter is placed between the last two solenoids to filter out positrons further.

The optics parameters applied in the transport beamline to optimize the transmission of the surface muon beam are summarized in Tab.~\ref{tab:optics}.

\begin{table*}[htbp]
\centering
\caption{The optics parameters applied in the extraction beam line to transport surface muon beam.}
\label{tab:optics}
\begin{tabular}{lcccc}
\toprule
 Components & Position (m) & Length (mm) & Aperture (mm) & Field (T) \\
\hline
Capture solenoid  & 0.47 & 373 & 500 & 0.432  \\
Dipole1  & 1.93 & 640 & 400 & -0.100 \\
Solenoid1  & 3.17 & 373 & 500 & 0.241  \\
Solenoid2  & 4.40 & 373 & 500 & 0.173  \\
Dipole2  & 6.05 & 640 & 400 & 0.100  \\
Solenoid3  & 7.17 & 373 & 500 & 0.136  \\
Solenoid4  & 8.00 & 373 & 500 & 0.199  \\
Dipole3  & 9.45 & 640 & 400 & -0.104 \\
Solenoid5  & 10.82 & 373 & 500 & 0.226  \\
Wien Filter   & 11.75 & 1500 & 300 &       \\
Solenoid6  & 13.18 & 373 & 500 & 0.440  \\
Exit & 13.58 &     &       \\
\hline
\end{tabular}
\end{table*}

\subsection{Optimization of the beamline optics}

To optimize the beamline optics, an algorithm integrating the pattern search method with the coordinate descent method was employed. The coordinate descent method optimizes a multivariate objective function by solving a series of univariate optimization problems sequentially. To facilitate the implementation, a cyclic approach was adopted, where each univariate problem is solved in turn, and once a complete cycle over all variables is finished, the process restarts with the first variable. Each univariate problem refines the solution estimate by optimizing selected variables while holding the others constant.

A pattern search method was chosen for solving the univariate optimization problems. This method relies solely on function evaluations and does not require the gradient of the function, making it particularly suitable for functions with hard-to-compute or unknown derivatives. Additionally, it is straightforward to implement. However, it is prone to converging to local maxima, and its effectiveness depends heavily on the initial value settings. Details of the implementation can be found in the Appendix~\ref{sec:appen1}.

\subsection{Expected surface muon beam properties}

\begin{figure}[htbp]
\includegraphics[width=0.9\linewidth]{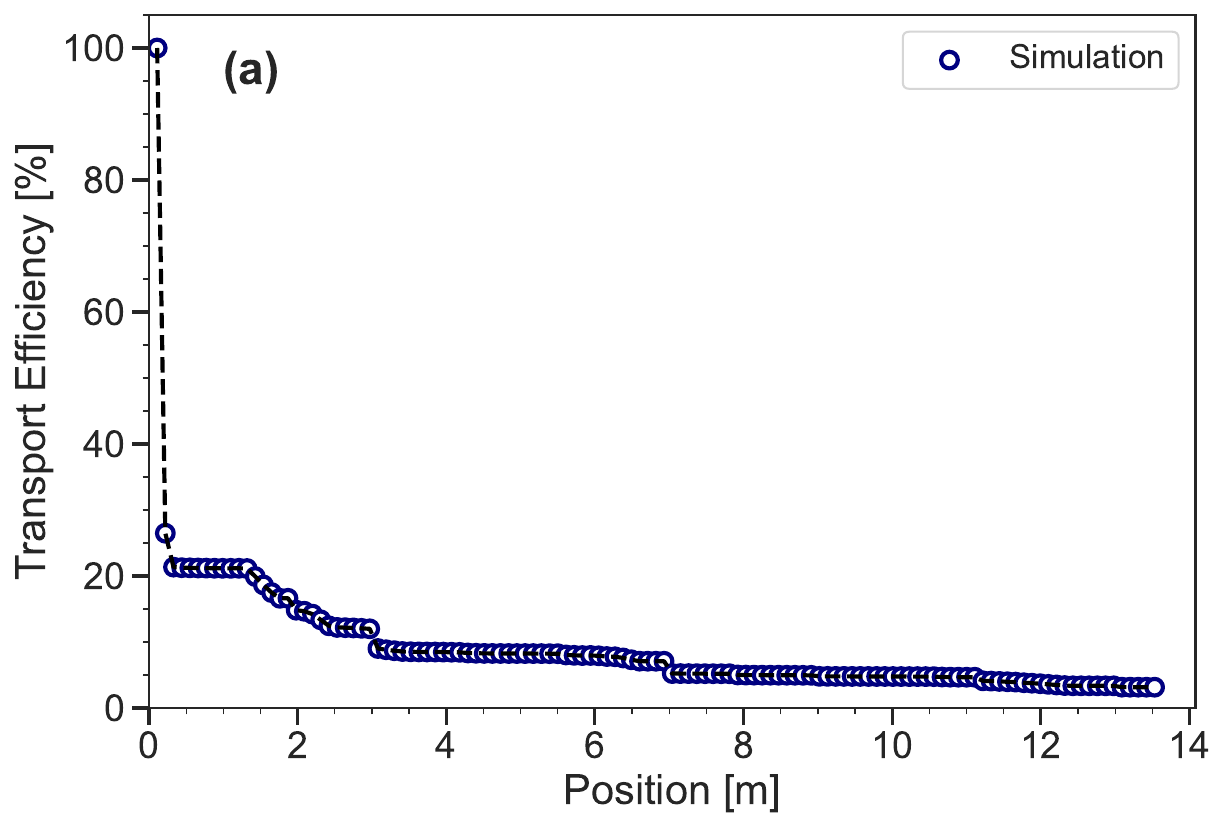}
\includegraphics[width=0.9\linewidth]{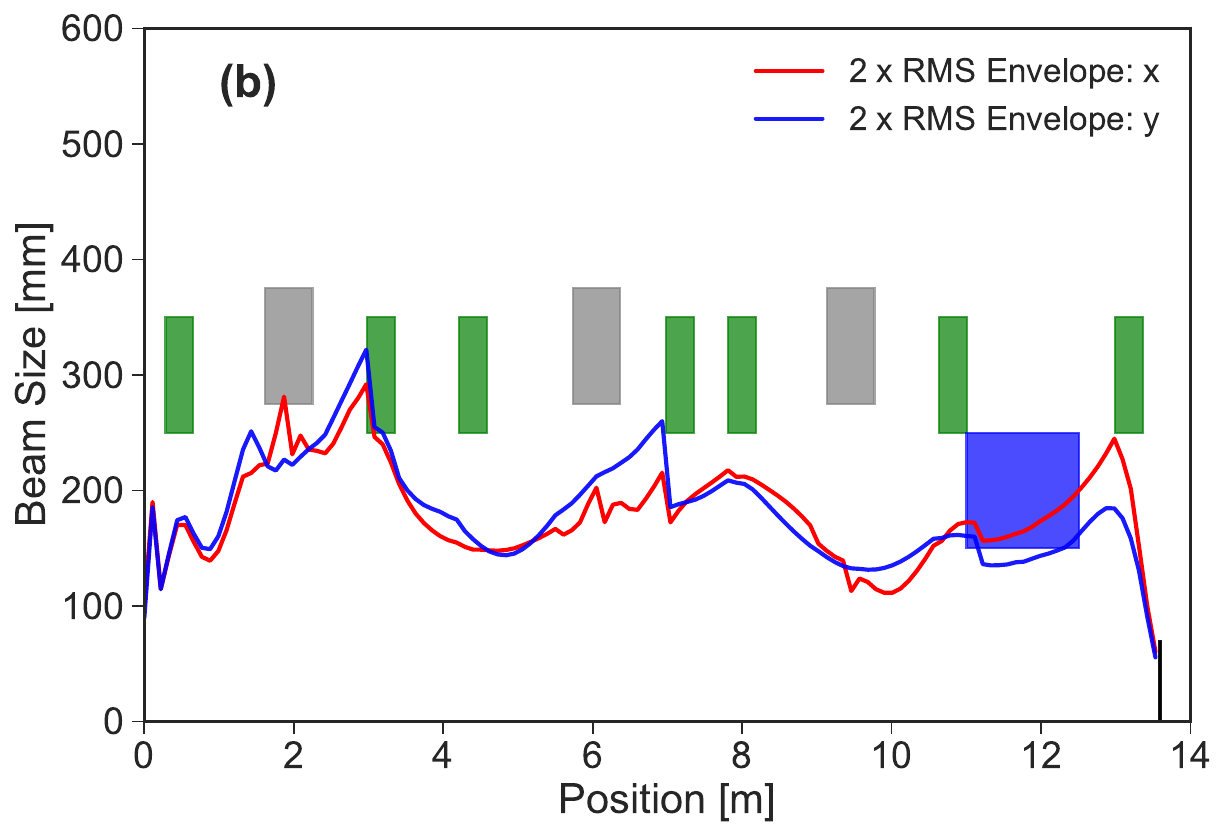}
\caption{(a) Transport efficiency along the surface muon beamline from the target to the experimental area. (b) Beam envelopes (2 $\times$ RMS) along the surface muon beamline from the target to the experimental area in G4Beamline simulation with optimized optics. The optics and their apertures are depicted. Green represents the solenoid magnets, gray indicates the bending magnets, and blue shows the Wien filter.}
\label{fig:transmission_envelope}
\end{figure}

\begin{figure*}[htbp]
\includegraphics[width=0.99\linewidth]{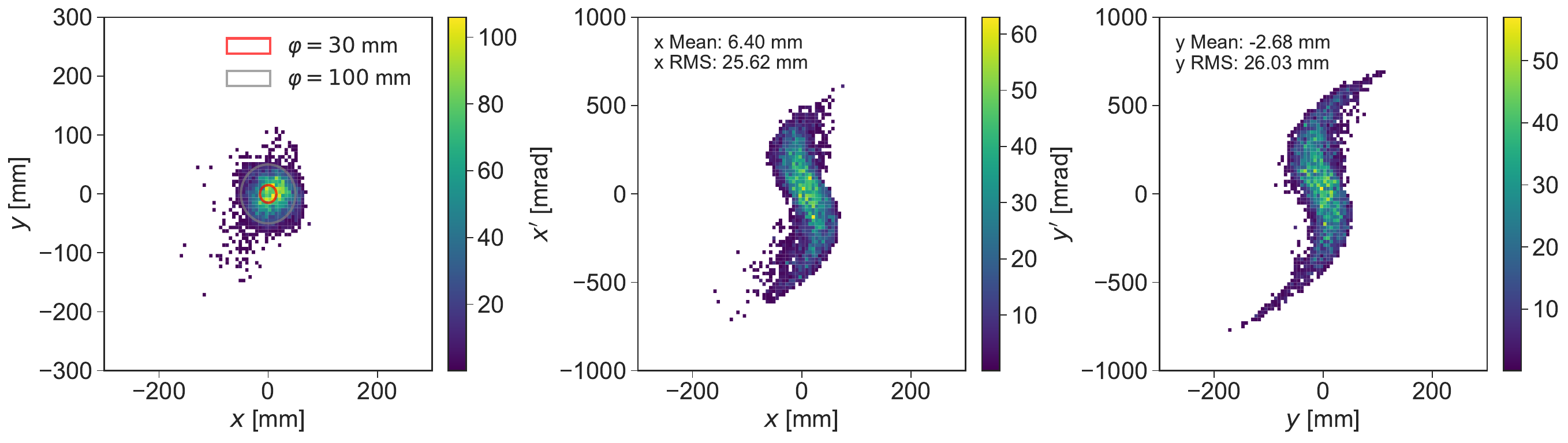}
\caption{The beam and phase-space distribution at the beam spot for all muons. From left to right are the $x-y$, $x-x'$, and $y-y'$ distributions. Color bars show the counts in the bins.}
\label{fig:beamspot_and_phasespace}
\end{figure*}

Figure~\ref{fig:transmission_envelope} illustrates typical simulated results for the beam envelope and transport efficiency of surface muons along the beamline. The transport efficiency for surface muons from the target to the experimental area, with standard beam optics tuning, is $3.1\% \pm 0.03\%$. This efficiency includes a capture efficiency of approximately $21.2\% \pm 0.1\%$ for surface muons in the capture solenoid and a transmission efficiency of about $14.8\% \pm 0.03\%$ along the transport beamline. The anticipated surface muon intensity at the final focus is 60 $\mu^{+}$/bunch or $3.1 \times 10^{6}\,\mu^{+}$/s (for a repetition rate of 50\,kHz), with an RMS beam size of 25.6\,mm (horizontal) $\times$ 26.0\,mm (vertical), a mean momentum of 27.6\,MeV/c with a momentum bite ($\Delta$p/p) of 3.69\% and a polarization of 88\%. The spin is rotated due to the Wien filter, and the 85\% is relative to the beamline axis; this could be further improved with more detailed studies on the Wien filter. The beam distribution and phase space at this beam location are illustrated in Fig.~\ref{fig:beamspot_and_phasespace}. Given that the sample size in a typical $\mu$SR experiment is around 30\,mm in diameter~\cite{Pak:2021, Zhou:2021tyb}, the distribution of surface muons within this range is particularly important. The muon intensity in this region is $4.7 \times 10^{5}\,\mu^{+}$/s.

\begin{figure}[htbp]
\includegraphics[width=0.9\linewidth,clip]{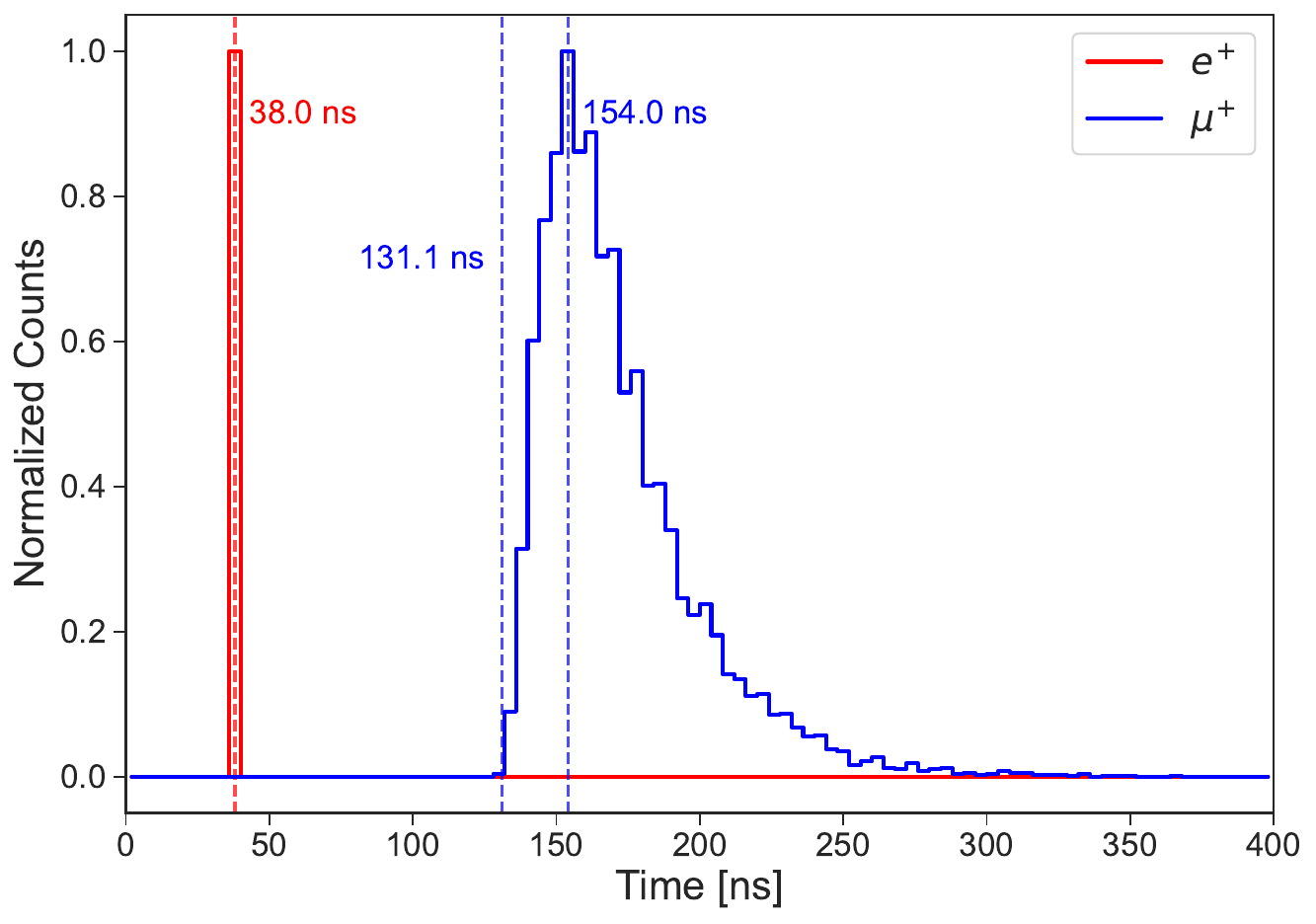}
\caption{Simulated time-of-flight distribution comparing positrons and muons 11 meters downstream from the target. Each histogram is normalized to its peak value. A distinct separation of approximately 100\,ns exists between the positron peak and the leading edge of the subsequent muon distribution.}
\label{fig:time_of_flight}
\end{figure}

The electron beam has a short pulse length of approximately 40\,fs~\cite{Gu:2024qww}. The muon arrival time distribution shows a Gaussian core with an exponential tail, resulting in a 20\,ns FWHM and extending beyond 200\,ns, which could affect timing resolution. Although detectors have an intrinsic time resolution of 100-200\,ps, the beam time spread, as shown in Fig.~\ref{fig:time_of_flight}, remains the main limiting factor in pulsed $\mu$SR. The mean parameters of the surface muon beam spot are summarized in Tab.~\ref{tab:beamline_para}.

\begin{table}[htbp]
\centering
\caption{Parameters of the surface muon beam spot at the experimental area, assuming a 50\,kHz operation for the pulsed electron beam. For polarization, the value just before the Wien filter is also noted.}
\label{tab:beamline_para}
\begin{tabular}{cc}
\toprule
 Parameters & Value\\
\hline
x/x' (RMS)  & 25.6\,mm / 221\,mrad   \\
y/y' (RMS)  & 26.0\,mm / 246\,mrad  \\
Mean momentum & 27.6\,MeV/c \\
$\Delta$p/p & ~3.69\%   \\
Polarization (before Wien Filter) &  97\%   \\
Polarization (after Wien Filter)  &  88\%   \\
$\mu^{+}$ rate (All) &  $3.1 \times 10^{6}\,\mu^{+}$/s \\
$\mu^{+}$ rate ($\phi$30\,mm) &  $4.7 \times 10^{5}\,\mu^{+}$/s \\
\hline
\end{tabular}
\end{table}

\subsection{Positron removal schemes}
In our target, approximately $1 \times 10^{10}$ positrons per bunch are generated. Some of these may contaminate the experimental area. Since positrons serve as the signal in $\mu$SR experiments, this presents a major background issue. To reduce positron contamination, we implemented a system that combines multiple bending magnets with a Wien filter. Parameters such as voltage and plate gap were chosen to be reasonable values with reference to the existing facility~\cite{Ikedo:2013}, and the plate length was determined considering the available length. The main parameters of the Wien filter are provided in Tab.~\ref{tab:wien_filter}.

\begin{table}[htbp]
\centering
\caption{Main parameters of the Wien filter}
\begin{tabular}{lc}
\hline
\textbf{Parameter} & \textbf{Value} \\
\hline
Electric field ($E_y$) & -2.67\,MV/m \\
Magnetic field ($B_x$) & 0.0328\,T \\
Length ($L$) & 1,500\,mm \\
Plate Gap ($g$) & 300\,mm \\
\hline
\end{tabular}
\label{tab:wien_filter}
\end{table}

For the positron removal study, we simulated $1 \times 10^{7}$ electron-on-target events. This provided sufficient statistics since the positron yield is very high. We detected 62 positrons at the $\phi$30\,mm beam spot following their transport through the beamline, which has been optimized for the surface muon. Scaling this result to SHINE's design electron bunch charge ($6.25 \times 10^{8}$ electrons), we projected that approximately 3,880 positrons would reach the beam spot per bunch while only around 10 muons can be transported. This calculation assumes linear scaling. Although this positron-to-muon ratio significantly decreases from the initial $10^{6}$ near the target, it still presents a major challenge for $\mu$SR experiments, which usually require positron contamination to be below 1\%.

Therefore, in addition to the Wien filter, we are exploring an alternative approach using a fast kicker with strip lines and a septum magnet to eliminate positrons from the beamline. The positron background in the beamline has the same momentum as the muon but experiences a different time of flight due to its varying mass. This difference could be utilized to remove positrons. As shown in Fig.~\ref{fig:time_of_flight}, the time-of-flight difference between positrons and muons at 11\,m from the target in this beamline is approximately 100\,ns.

Advanced XFEL facilities utilize fast kicker systems to distribute high-energy electron beams across multiple beamlines at MHz-level repetition rates. For example, at SHINE, a lumped inductance kicker is currently being developed utilizing a vacuum chamber equipped with a single-turn coil~\cite{Liu2024}. This kicker is capable of operating at a frequency of 1 MHz, possesses a magnetic pulse width of 600\,ns, and achieves a full magnetic strength of 0.005\,T; at DESY's European XFEL, 20\,GeV electron beams operate at 4.5 MHz, with excess electrons not used for lasing being dumped by fast kickers that have maximum pulse widths of 30\,ns and rise/fall times of 15\,ns~\cite{Obier:2019fel}. By designing a similar fast kicker system specifically for muon beamlines, we believe the positron contamination issue can be effectively resolved. This promising approach, along with design optimization studies and comprehensive performance analysis, should be thoroughly explored in an upcoming publication dedicated to positron removal for muon experiments.

\section{Conclusions}
This article introduced the concept of a muon source utilizing a high-repetition-rate electron beam and a transport beamline design. A muon beam with a repetition rate of approximately 50\,kHz, optimal for typical muon experiments, can be generated. Compared to current proton-driven muon sources, the relatively low muon production rate is compensated by the high repetition rate, resulting in a muon yield comparable to existing sources.

Using a straightforward beamline configuration consisting of solenoids, bending magnets, and a Wien filter, the surface muon intensity delivered to the experimental area was estimated to be approximately $4.7 \times 10^{5}\,\mu^{+}$/s for a beam spot $\phi$30 mm. The realization of this concept requires further detailed design studies, including optimization of target geometry, thermo-mechanical calculations for the target, and proof-of-principle experiments. The removal of positrons is particularly challenging. Once constructed, this high-repetition-rate pulsed muon beamline is well-suited for applications in $\mu$SR spectrometer~\cite{Li:2023gxn}, muon electric dipole moment (EDM) searches~\cite{Adelmann:2021udj,Adelmann:2025nev}, and muonium-to-antimuonium conversion experiments~\cite{Han:2021nod,Bai:2022sxq,Bai:2024skk}. Additionally, its unique timing structure complements the Chinese Muon Facilities currently under construction at the Chinese Spallation Neutron Source (CSNS)~\cite{Bao:2023nup} in Dongguan, Guangdong, China, and is being studied with the Initiative Accelerator Driven Subcritical System (CiADS)~\cite{Cai:2023caf} and the High-Intensity Heavy-Ion Accelerator Facility (HIAF)~\cite{Xu:2025spd} in Huizhou, Guangdong, China. 

\begin{acknowledgments}
We are extremely grateful to Yu Bao, Vadim Grinenko, Hong Ding, Bo Liu, Zhi Liu, Jian Tang, Dao Xiang, and Zhentang Zhao for fruitful discussions regarding the beamline optimization and physics potential of the SHINE muon beam. F.~Liu, S.~Chen, and L.~Wang were supported by the Participation in Research Program (PRP) of Shanghai Jiao Tong University (Program number T426PRP43001). The work of M.~Lyu, Y.~Takeuchi, J.~Wang, and K.~S.~Khaw was supported by the Shanghai Pilot Program for Basic Research (Grant number 21TQ1400221).
\end{acknowledgments}

\section*{Appendix 1: Muon Yield Comparison between G4Beamline and FLUKA}
\label{sec:appen1}

\begin{figure}[htbp]
\includegraphics[width=0.7\linewidth,clip]{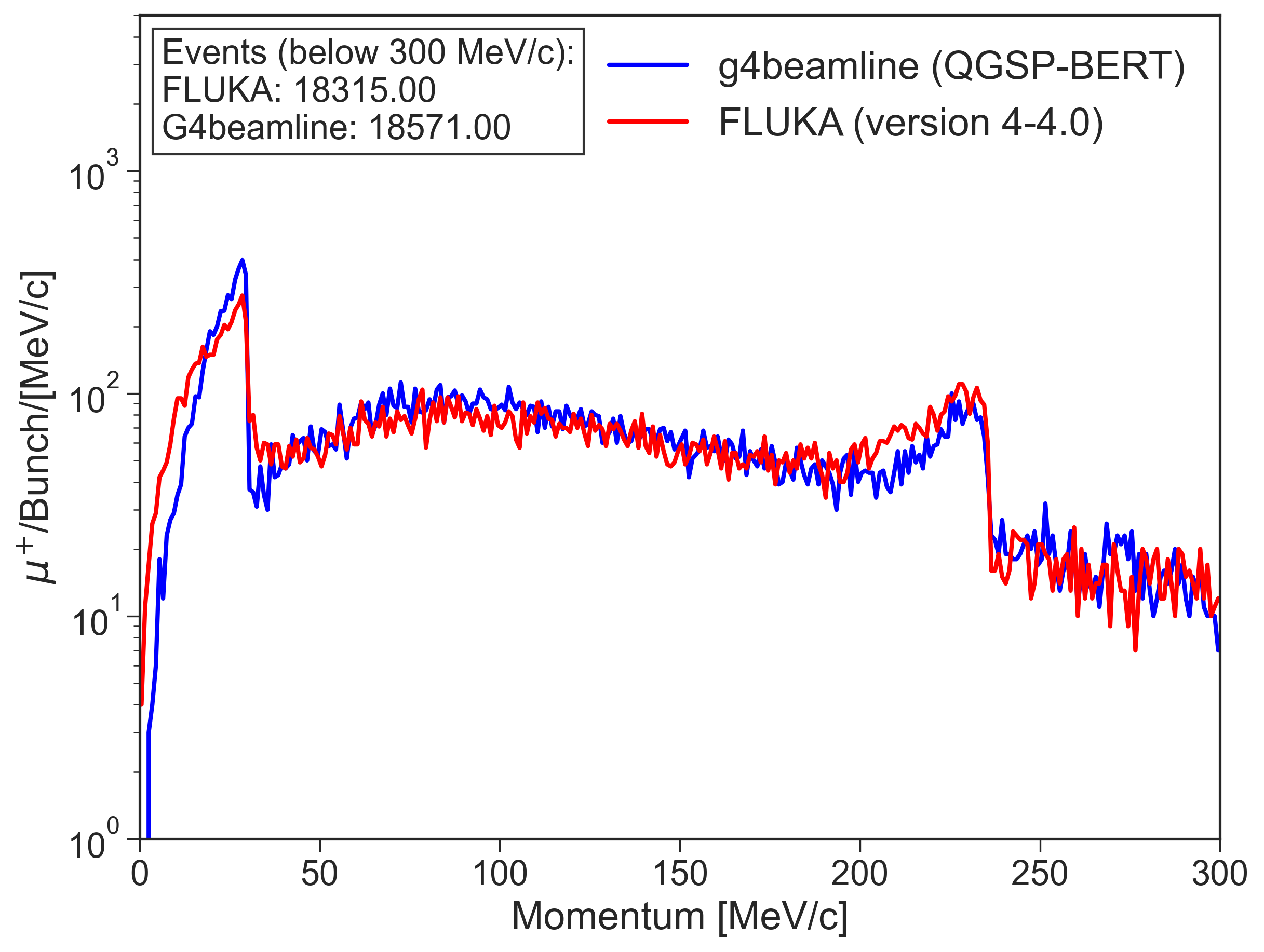}
\caption{A comparison of positive muon yield between G4Beamline and FLUKA.}
\label{fig:g4bl_fluka_comparison}
\end{figure}

Figure~\ref{fig:g4bl_fluka_comparison} shows a comparison of positive muon yield between G4Beamline and FLUKA.
No significant differences are observed in the yield for momentum regions below 300 MeV/c, showing generally good agreement between the two calculation codes. However, there are some discrepancies in behavior near the two peaks. In the momentum region of 25-30 MeV/c, the yields differ by up to approximately 30\%. The accuracy of these simulation models needs to be quantitatively evaluated through future experimental validation.

\section*{Appendix 2: Pattern search algorithm for the beamline optimization}
\label{sec:appen2}
This algorithm ensures a balanced search throughout the variables in a multi-dimensional optimization problem. An example of the procedure for the algorithm is described below. In this example, we consider maximizing the value of the multivariate function $f(x_{1},x_{2}, ... ,x_{n})$, where $ i = 1, 2, ..., n $, and $ n $ is the number of variables.
\begin{enumerate}
\item Initialization:
    \begin{itemize}
    \item Set the initial search points $ x_{i} $  and initial step sizes $ \Delta_{i} $.
    \end{itemize}
\item Iteration:
    \begin{itemize}
    \item For each variable $ x_{i} $ and step size $ \Delta_{i} $, proceed with the following steps.
    \item Evaluate the function at three points $ x_{i} - \Delta_{i} $, $ x_{i} $, $ x_{i} + \Delta_{i} $.
    \item Select the point with the highest function value and update the search point $ x_{i} $ to this point if it is different from the current search point.
    \item Then, adjust the step size $ \Delta_{i} $; maintain if there was an update of the search point, halves it if not.
    \end{itemize}
\item Update:
    \begin{itemize}
    \item After completing a cycle of searches across all variables, return to the first variable and repeat the search process from the updated search point $x_{i} $.
    \end{itemize}
\item Termination:
    \begin{itemize}
    \item Terminate the algorithm if the step size falls below a pre-defined threshold.
    \end{itemize}
\end{enumerate}

Here, the objective function was the number of surface muons transported to the experimental area, the variables were the position and field strength of each optics component. To avoid local maxima as much as possible, simulations were initially performed using several randomly generated beamline configurations, and the configuration with the highest number of surface muon transported was used as the initial search point.

\bibliography{bibliography}

\end{document}